\documentclass[conference]{IEEEtran}

\IEEEoverridecommandlockouts

\usepackage{amsmath,amssymb}
\usepackage{graphicx}
\usepackage{booktabs}
\usepackage{multirow}
\usepackage{url}
\usepackage{cite}
\usepackage[hidelinks,hypertexnames=false]{hyperref}
\usepackage{algorithm}
\usepackage{algorithmic}
\usepackage{xcolor}
\usepackage{balance}
\usepackage{placeins}
\usepackage{float}

\title{FastHeapDict: Permanent Slot Indexing for Efficient Decrease-Key Operations in Indexed Priority Queues}

\author{
\IEEEauthorblockN{Hojjat T. Goudarzi}
\IEEEauthorblockA{
Oregon State University\\
Corvallis, Oregon, USA\\
torabigh@oregonstate.edu
}
\and
\IEEEauthorblockN{Arash Torabi Goodarzi}
\IEEEauthorblockA{
University of Würzburg\\
Würzburg, Germany\\
arash.goodarzi@uni-wuerzburg.de
}
\and
\IEEEauthorblockN{Mohammad Amir Eshraghi}
\IEEEauthorblockA{
Purdue University\\
West Lafayette, Indiana, USA\\
meshragh@purdue.edu
}
}

\begin{document}
\maketitle

\begin{abstract}
Efficient decrease-key support remains an important practical concern in priority queues used by shortest-path and graph-processing algorithms. Although duplicate-entry heaps avoid explicit decrease-key operations and are simple to implement, they may accumulate many stale entries when tentative priorities are repeatedly improved. This paper presents \emph{FastHeapDict}, an indexed binary min-heap based on \emph{permanent slot indexing}. Instead of repeatedly locating keys through hash-table lookups during heap repair, FastHeapDict assigns each key a stable integer slot and stores that slot directly in heap entries. As a result, position updates during sift-up and sift-down operations are performed through integer indexing while preserving the usual indexed-heap interface.

We evaluate FastHeapDict in both Python and C++20 using Dijkstra's algorithm as the primary experimental workload. The benchmarks include constructed dense stress graphs, randomized dense graph variants, and comparisons against duplicate-entry heaps and existing priority-queue alternatives. Across the tested Python workloads, FastHeapDict consistently outperforms the duplicate-entry \texttt{heapq} baseline and the tested \texttt{heapdict} and \texttt{pqdict} implementations. In the C++20 implementation, FastHeapDict also achieves strong constant-factor improvements over \texttt{std::priority\_queue} duplicates, \texttt{std::set}-based decrease-key, and \texttt{std::multiset} duplicates. These results indicate that permanent slot indexing can provide a compact and effective implementation strategy for indexed priority queues in workloads where priority updates are frequent.
\end{abstract}

\section{Introduction}

Priority queues are fundamental components of many graph algorithms and optimization problems. Among their most important operations is \emph{decrease-key}, which allows the priority associated with an existing element to be updated efficiently without inserting duplicate entries. Algorithms such as Dijkstra's single-source shortest-path algorithm rely heavily on this operation when tentative distances are repeatedly improved during execution \cite{dijkstra1959note}.

Despite the theoretical importance of decrease-key, many practical implementations of Dijkstra's algorithm avoid supporting it explicitly. Instead, they rely on a duplicate-entry strategy in which every distance improvement inserts a new heap entry while obsolete entries are discarded only when they are eventually removed from the priority queue. This approach is simple, robust, and readily supported by widely used libraries such as Python's \texttt{heapq} \cite{pythonheapq} and C++'s \texttt{std::priority\_queue} \cite{cppreferencepriorityqueue}. However, when many priority updates occur before vertices are finalized, duplicate entries accumulate, increasing the number of heap operations, memory usage, and unnecessary processing of stale elements.

Indexed priority queues address this limitation \cite{cormen2009introduction} by maintaining exactly one live entry for each key and updating its priority in place. Traditional indexed heaps, however, typically maintain a mapping from keys to heap positions, requiring repeated hash-table or dictionary lookups whenever heap elements move during sift-up or sift-down operations. Although these lookups have expected constant-time complexity, they contribute additional constant-factor overhead in implementations where decrease-key operations occur frequently.

This paper introduces \emph{FastHeapDict}, an indexed binary min-heap based on the idea of \emph{permanent slot indexing}.

Unlike conventional indexed heaps that repeatedly recover heap positions through key-based hash-table accesses, FastHeapDict performs all subsequent heap maintenance through direct integer indexing once a permanent slot has been assigned. Each key is assigned a stable integer slot when it first appears, and heap entries store this slot together with the key and its priority. A separate position array maps slots to current heap locations, allowing heap repair operations to update positions through integer indexing rather than repeated key-based hash lookups. The resulting implementation preserves the simplicity and asymptotic complexity of a conventional indexed binary heap while reducing practical overhead during decrease-key operations.

The contribution of this work is implementation-oriented rather than asymptotic. FastHeapDict does not change the theoretical complexity of indexed binary heaps; instead, it proposes a compact implementation strategy that reduces repeated hash-table access during heap maintenance. To evaluate its practical effectiveness, we implement FastHeapDict independently in both Python and C++20 and use Dijkstra's algorithm as the primary evaluation workload. The proposed design is compared against duplicate-entry heaps as well as existing indexed and tree-based priority-queue implementations across several controlled benchmark families.

The remainder of this paper is organized as follows. Section~\ref{sec:related} reviews existing priority-queue designs and practical implementations of Dijkstra's algorithm. Section~\ref{sec:fastheapdict} presents the design of FastHeapDict. Section~\ref{sec:cpp} describes the C++20 implementation. Section~\ref{sec:methodology} outlines the experimental methodology. Sections~\ref{sec:python-results} and~\ref{sec:cpp-results} present the Python and C++20 experimental results, respectively. Sections~\ref{sec:discussion} and~\ref{sec:limitations} discuss the findings and limitations of the study, and Section~\ref{sec:conclusion} concludes the paper.

\section{Related Work}
\label{sec:related}

Priority queues have been studied extensively because they are fundamental components of graph algorithms, scheduling, simulation, and many other optimization problems. Among practical implementations, the binary heap remains the most widely used choice because of its simplicity, low memory overhead, and predictable logarithmic-time operations \cite{williams1964heapsort,cormen2009introduction}. When combined with an indexed decrease-key operation, a binary heap supports the standard $O((V+E)\log V)$ implementation of Dijkstra's shortest-path algorithm. In contrast, implementations that avoid decrease-key by inserting duplicate entries may perform substantially more heap operations, since every successful relaxation creates a new queue element while obsolete entries remain until they are eventually removed.

Considerable research has explored priority-queue structures with improved theoretical performance. Fibonacci heaps achieve constant amortized time for decrease-key and lead to stronger asymptotic bounds for several graph algorithms \cite{fredman1987fibonacci}. Pairing heaps were later proposed as a simpler self-adjusting alternative that often performs well in practice while retaining many of the advantages of more sophisticated heap structures \cite{fredman1986pairing}. Despite these theoretical developments, binary heaps continue to dominate many production implementations because their lower constant factors and implementation simplicity frequently outweigh asymptotic improvements for practical workloads.

Many widely used software libraries expose priority queues without providing explicit decrease-key support. Python's \texttt{heapq} module implements an efficient binary heap but leaves indexed priority updates to the application \cite{pythonheapq}. Consequently, Dijkstra's algorithm is often implemented using duplicate entries together with lazy deletion of stale elements. Third-party libraries such as \texttt{pqdict} \cite{pqdictdocs} and \texttt{heapdict} \cite{heapdictgithub} provide indexed priority queues with direct priority updates. Similarly, the C++ Standard Library offers \texttt{std::priority\_queue} \cite{cppreferencepriorityqueue}, which follows the duplicate-entry approach in practical shortest-path implementations, while ordered containers such as \texttt{std::set} \cite{cppreferenceset} and \texttt{std::multiset} \cite{cppreferencemultiset} can be adapted to support priority updates through erase-and-reinsert operations.

FastHeapDict belongs to the family of indexed binary heaps, but differs in how the mapping between keys and heap positions is maintained. Conventional indexed implementations typically update a key-to-position map whenever heap elements move during sift-up or sift-down operations. FastHeapDict instead assigns each key a permanent integer slot and records heap positions using slot indices rather than repeatedly accessing the original keys. The motivation is not to change the asymptotic complexity of indexed heaps, but to reduce practical constant-factor overhead associated with repeated key hashing during heap maintenance. The following sections describe this design in detail and evaluate its effectiveness using Dijkstra's algorithm as the primary benchmark workload.

\section{FastHeapDict Design}
\label{sec:fastheapdict}

\subsection{Design Motivation}

The goal of FastHeapDict is not to change the theoretical complexity of indexed binary heaps, but to reduce the implementation overhead incurred during practical decrease-key workloads.

Indexed binary heaps support efficient \emph{decrease-key} operations \cite{cormen2009introduction} by maintaining a mapping from each key to its current position in the heap. A conventional implementation stores this mapping directly as a key-to-index dictionary. Whenever two heap entries are exchanged during \textsc{BubbleUp} or \textsc{BubbleDown}, the corresponding dictionary entries must also be updated. Although each lookup or update has expected constant-time complexity, repeated key hashing becomes part of every heap repair operation. In workloads that perform frequent priority updates, this overhead contributes to the overall constant factors of the implementation.

FastHeapDict addresses this issue by separating key identification from heap maintenance. Rather than storing heap positions directly in the hash table, each key is assigned a permanent integer slot when it first appears. Heap entries carry this slot together with the key and its priority, while a separate position array maps slots to the current heap indices. Once the slot has been obtained, all subsequent heap maintenance is performed through integer indexing without further key-based hash lookups. The resulting design preserves the behavior and asymptotic complexity of a conventional indexed binary heap while reducing the amount of hash-table activity during heap movement.

\paragraph{Definition (Permanent Slot Indexing).} Permanent slot indexing is a position-management strategy in which each key is assigned a unique integer slot exactly once during its lifetime. Heap entries retain this slot permanently, while a separate position array maps slots to current heap locations. Consequently, heap maintenance relies on direct integer indexing instead of repeated key-based hash lookups.

\subsection{Data Structures}

FastHeapDict is organized around three complementary data structures:

\begin{itemize}
    \item \texttt{slot}: a hash table mapping each key to its permanent slot;
    \item \texttt{pos}: an integer array mapping each slot to the current heap position, or $-1$ if the key is not currently present in the heap;
    \item \texttt{heap}: a binary min-heap, following the array-based heap representation introduced for heapsort \cite{williams1964heapsort}, whose entries store the triple $(priority,\; key,\; slot)$.
\end{itemize}

Each key receives its slot only once. The slot remains permanently associated with that key, even after the corresponding heap entry has been removed. If the key is inserted again later, the existing slot is reused. During Dijkstra's algorithm each vertex normally has at most one live heap entry, but permanent slot indexing also provides a consistent representation for more general insertion and reinsertion scenarios.

\subsection{Priority Updates}

Insertion and priority updates begin with a single lookup in the \texttt{slot} table. If the key has not appeared previously, a new slot is allocated, appended to the position array, and stored together with the newly created heap entry before restoring the heap property using \textsc{BubbleUp}. If the key already has a live heap entry, its priority is modified in place and the heap is repaired by invoking either \textsc{BubbleUp} or \textsc{BubbleDown}, depending on whether the priority decreases or increases.

Heap repair is implemented using the \emph{hole method}. Rather than repeatedly swapping adjacent heap entries, the updated element is temporarily removed while parent or child entries are shifted into the vacant position. Since every heap entry carries its permanent slot identifier, the corresponding position array can be updated immediately through integer indexing without consulting the key-to-slot hash table. This organization reduces repeated hashing while preserving the standard behavior of binary heap operations.

Algorithm~\ref{alg:insert-update} summarizes the insert-or-update procedure.

\begin{algorithm}[t]
\caption{FastHeapDict insert-or-update}
\label{alg:insert-update}
\begin{algorithmic}[1]
\REQUIRE key $k$, priority $p$
\STATE $s \leftarrow$ slot assigned to $k$, creating one if needed
\IF{$s$ is new or $pos[s] < 0$}
    \STATE append $(p,k,s)$ to heap
    \STATE $pos[s] \leftarrow$ index of new entry
    \STATE \textsc{BubbleUp} from $pos[s]$
\ELSE
    \STATE $i \leftarrow pos[s]$
    \STATE $old \leftarrow heap[i].priority$
    \STATE $heap[i].priority \leftarrow p$
    \IF{$p < old$}
        \STATE \textsc{BubbleUp} from $i$
    \ELSIF{$p > old$}
        \STATE \textsc{BubbleDown} from $i$
    \ENDIF
\ENDIF
\end{algorithmic}
\end{algorithm}

\subsection{Complexity Analysis}

Let $V$ denote the number of vertices and $E$ the number of edges. In the indexed implementation of Dijkstra's algorithm, FastHeapDict maintains at most one live heap entry for each vertex. Insertions, priority updates, and minimum removals therefore require $O(\log V)$ heap operations together with an expected $O(1)$ key-to-slot lookup at the beginning of each insertion or update. Consequently, Dijkstra's algorithm retains its standard running time of $O((V+E)\log V)$ while requiring $O(V)$ auxiliary space for the priority queue. Including the graph representation, the overall space complexity remains $O(V+E)$.

The duplicate-entry strategy exhibits a different practical behavior. Every successful edge relaxation may insert an additional heap entry, allowing obsolete entries to accumulate until they are eventually removed. In the worst case, this results in up to $O(E)$ heap entries and corresponding stale removals, increasing both memory consumption and the amount of heap maintenance performed during execution. The experimental evaluation presented later is designed to highlight these practical differences while recognizing that both approaches retain the same asymptotic running time.
\section{C++20 Implementation}
\label{sec:cpp}

The C++20 implementation follows the same algorithmic design as the Python version while leveraging static typing and standard-library containers \cite{cppreferenceunorderedmap,cppreferencevector}. The permanent slot indexing strategy is preserved without modification, allowing the experimental comparison to focus on implementation language and runtime characteristics rather than differences in data structures or algorithms.

The key-to-slot mapping is implemented using \texttt{std::unordered\_map<Key, std::size\_t>} \cite{cppreferenceunorderedmap}. Heap positions are stored in a \texttt{std::vector<int>} \cite{cppreferencevector}, where a value of $-1$ indicates that the corresponding key is not currently present in the heap. The binary heap itself is maintained as a \texttt{std::vector<Node>}, with each node storing a priority, a key, and its permanent slot identifier.

Insertion and update operations use \texttt{try\_emplace} \cite{cppreferenceunorderedmap} to combine slot lookup and allocation into a single operation. When a key is inserted for the first time, a new heap node is appended and restored to the correct position using \textsc{BubbleUp}. If the key already has a live heap entry, its priority is updated in place, after which either \textsc{BubbleUp} or \textsc{BubbleDown} is performed depending on whether the priority decreases or increases.

As in the Python implementation, heap repair uses the hole method rather than repeated adjacent swaps. During heap movement, the position array is updated directly through the permanent slot stored in each node. Consequently, heap maintenance never performs additional lookups in the key-to-slot hash table after the initial slot has been obtained.

To evaluate the practical behavior of the proposed design, the C++ benchmark compares four implementations: FastHeapDict, \texttt{std::priority\_queue} using duplicate entries, \texttt{std::set} using erase-and-reinsert updates as a tree-based decrease-key implementation \cite{cppreferenceset}, and \texttt{std::multiset} using duplicate entries \cite{cppreferencemultiset}. These baselines distinguish two independent design choices: maintaining a single live entry versus allowing duplicate entries, and using a binary heap versus an ordered tree as the underlying priority-queue structure.
\section{Experimental Methodology}
\label{sec:methodology}

\subsection{System Configuration}
\label{sec:system_configuration}
All experiments were performed on the same hardware and software platform to ensure a consistent evaluation environment. The benchmark system ran Fedora~44 \cite{fedoradocs} on a 12th Generation Intel Core i7-12800H processor with 64~GB of RAM. The C++ implementation was compiled using GCC~16.1.1 with CMake~4.3.0 and the C++20 language standard \cite{gccdocs,cmakedocs,cpp20standard}. The Python implementation was executed using Python~3.12 within a dedicated Conda environment \cite{python312docs,condadocs}.

\subsection{Timing Methodology}

Both the Python and C++20 implementations follow the same timing methodology. The input edge list is converted into an adjacency-list representation before timing begins. The measured interval therefore includes only priority-queue initialization and execution of Dijkstra's algorithm, excluding graph construction and preprocessing overhead.

For the Python implementation, cyclic garbage collection is temporarily disabled \cite{pythongc} only during each timed execution and restored immediately afterward to reduce measurement variability. The execution order of competing methods is rotated across trials to minimize ordering effects, and all reported running times correspond to the median over repeated executions.

The C++ benchmark records both a CSV file containing per-trial timing measurements and a terminal-style log summarizing the execution of every benchmark.

\subsection{Correctness Verification}

All benchmark results were obtained only after passing a comprehensive collection of correctness tests.

The Python implementation first performs a direct hashing validation, followed by 100 randomized priority-queue operation sequences and 500 randomized positive-weight Dijkstra comparisons between the indexed and duplicate-entry implementations.

The C++20 implementation performs analogous correctness checks for FastHeapDict together with randomized graph validation. Every benchmark reported in this paper passed all self-tests before any performance measurements were collected.

\subsection{Graph Families}

The evaluation uses undirected graphs with positive edge weights generated deterministically after fixing the random seed.

Experiments~1--3 employ a controlled dense stress-graph construction designed to emphasize workloads with frequent priority updates. Vertex~0 serves as the source. A set of intermediate \emph{improver} vertices repeatedly decreases the tentative distances of a collection of active sink vertices. The final target vertex is assigned a larger shortest-path distance than the remaining sinks, forcing duplicate-entry implementations to process many obsolete heap entries before termination. Additional high-weight filler edges are inserted until the desired graph density is reached; these edges increase density without affecting the intended shortest-path structure.

Graph density is measured relative to the complete undirected graph,

\begin{equation}
\rho = \frac{E}{V(V-1)/2},
\end{equation}

where $E$ and $V$ denote the numbers of edges and vertices, respectively. For completeness, the experiments also report

\begin{equation}
\alpha = \frac{E}{V^2},
\end{equation}

allowing dense graphs to be described using the relation $E=\alpha V^2$.

Experiment~1 varies the number of vertices while fixing the density ratio at $\rho=0.25$. Experiment~2 fixes $V=1500$ and varies graph density. Experiment~3 compares every runnable priority-queue implementation using $V=1500$ and $\rho=0.10$. Experiment~4 evaluates randomized dense stress graphs with $V=1500$ and $\rho=0.25$ over multiple random seeds. In this randomized benchmark, the active sink set, the number of active sinks, sink ordering, base distances, and filler edges all vary between trials while preserving the decrease-key-intensive characteristics of the workload.
\section{Python Results}
\label{sec:python-results}

\subsection{Scaling with the Number of Vertices}

\begin{figure}[t]
    \centering
    \includegraphics[width=\linewidth]{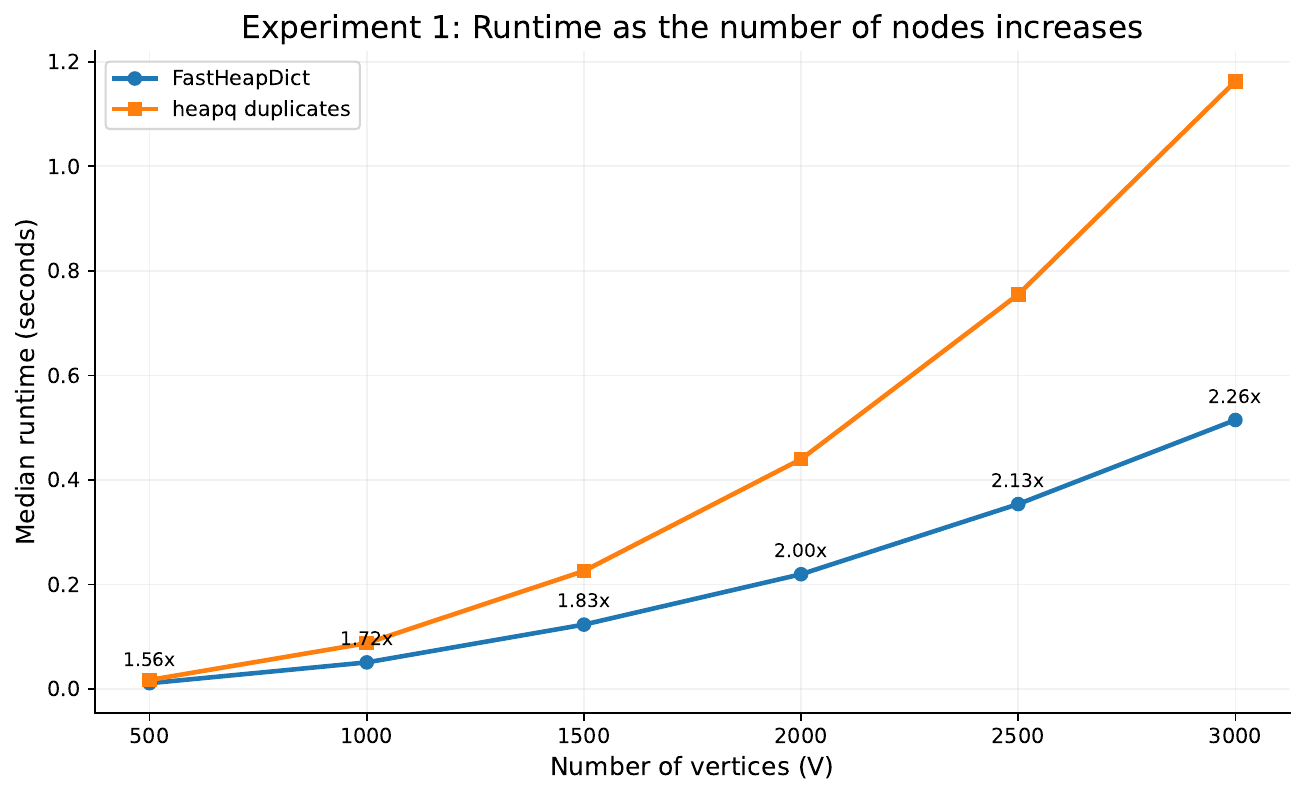}
    \caption{Python Experiment~1: running time as the number of vertices increases while the density ratio is fixed at $0.25$.}
    \label{fig:python_exp1}
\end{figure}

Table~\ref{tab:python_scaling} summarizes the median running times obtained in Experiment~1, and Fig.~\ref{fig:python_exp1} visualizes the corresponding runtime trend. The graph density ratio is fixed at $\rho=0.25$, while the number of vertices increases from 500 to 3000. Across the entire range, FastHeapDict consistently outperforms the duplicate-entry \texttt{heapq} implementation. The observed speedup increases from $\mathbf{1.56\times}$ for the smallest graph to $\mathbf{2.26\times}$ for the largest, indicating that the performance advantage becomes more pronounced as the workload grows.

\begin{table}[t]
\caption{Python median running time as the graph size increases at density ratio $0.25$.}
\label{tab:python_scaling}
\centering
\scriptsize
\begin{tabular}{l r r r r r}
\toprule
$V$ & $E$ & Density & FastHeapDict (s) & heapq dup. (s) & Speedup \\
\midrule
500   & 31,188    & 0.250 & 0.0110 & 0.0172 & 1.56x \\
1,000 & 124,875   & 0.250 & 0.0508 & 0.0876 & 1.72x \\
1,500 & 281,062   & 0.250 & 0.1231 & 0.2257 & 1.83x \\
2,000 & 499,750   & 0.250 & 0.2195 & 0.4396 & 2.00x \\
2,500 & 780,938   & 0.250 & 0.3538 & 0.7548 & 2.13x \\
3,000 & 1,124,625 & 0.250 & 0.5146 & 1.1625 & \textbf{2.26x} \\
\bottomrule
\end{tabular}
\end{table}

\subsection{Changing Graph Density}

\begin{figure}[t]
    \centering
    \includegraphics[width=\linewidth]{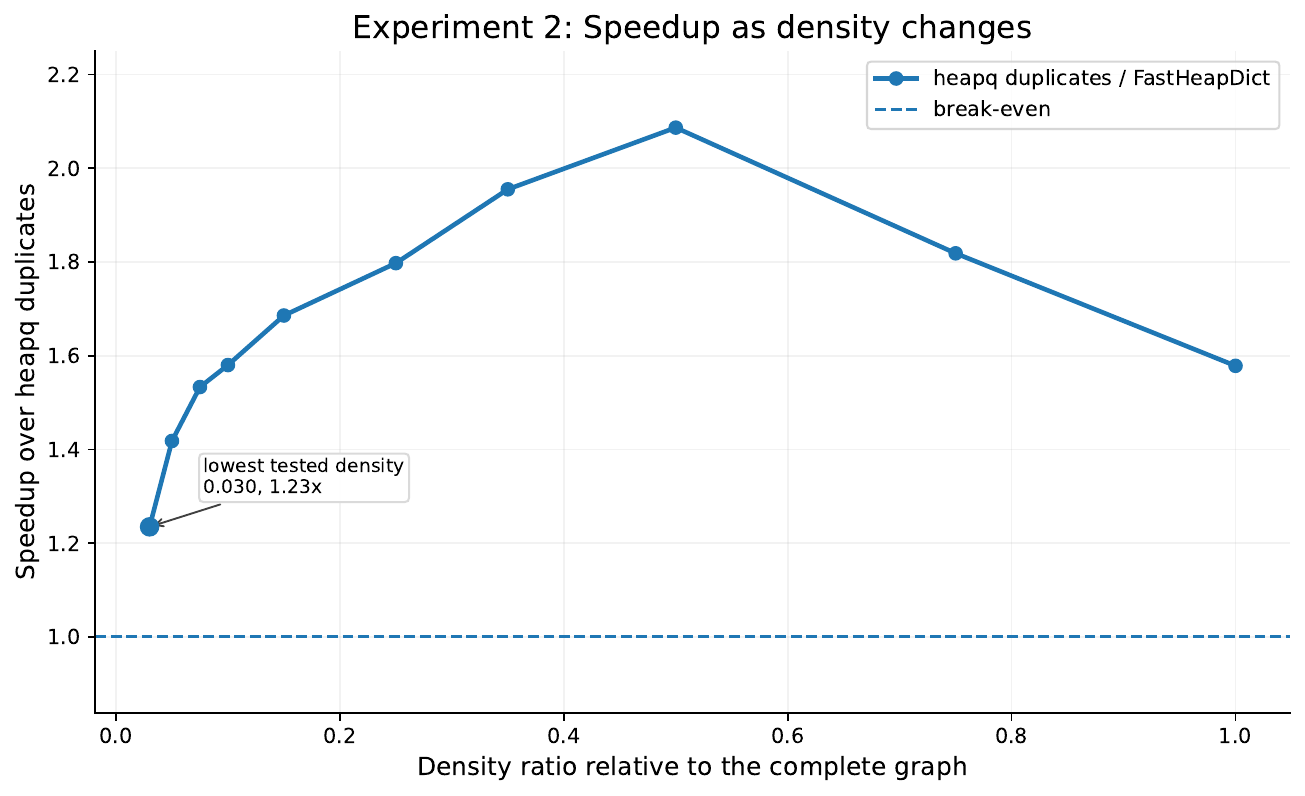}
    \caption{Python Experiment~2: speedup of FastHeapDict relative to duplicate-entry \texttt{heapq} as graph density varies.}
    \label{fig:python_exp2}
\end{figure}

Experiment~2 isolates the effect of graph density by fixing the number of vertices at $V=1500$ and varying the density ratio from $0.03$ to $1.0$. Table~\ref{tab:python_density} reports the corresponding median running times, and Fig.~\ref{fig:python_exp2} shows the resulting speedup trend over the density sweep. FastHeapDict remains consistently faster than the duplicate-entry baseline across all tested densities. Measured speedups range from $\mathbf{1.23\times}$ to $\mathbf{2.09\times}$, with the largest speedup occurring at density ratio $0.50$. As expected, the performance gap narrows for the sparsest graphs, where fewer repeated priority updates occur and duplicate entries accumulate less frequently.

\begin{table}[t]
\caption{Python median running time as graph density varies at $V=1500$.}
\label{tab:python_density}
\centering
\scriptsize
\setlength{\tabcolsep}{3.0pt}
\begin{tabular}{l r r r r r}
\toprule
Density & $E$ & $\alpha$ & FastHeapDict (s) & heapq dup. (s) & Speedup \\
\midrule
1.000 & 1,124,250 & 0.4997 & 0.5250 & 0.8287 & 1.58x \\
0.750 & 843,188   & 0.3748 & 0.3783 & 0.6880 & 1.82x \\
0.500 & 562,125   & 0.2498 & 0.2544 & 0.5309 & \textbf{2.09x} \\
0.350 & 393,488   & 0.1749 & 0.1795 & 0.3509 & 1.96x \\
0.250 & 281,062   & 0.1249 & 0.1283 & 0.2306 & 1.80x \\
0.150 & 168,638   & 0.0750 & 0.0755 & 0.1272 & 1.69x \\
0.100 & 112,425   & 0.0500 & 0.0491 & 0.0775 & 1.58x \\
0.075 & 84,319    & 0.0375 & 0.0355 & 0.0544 & 1.53x \\
0.050 & 56,212    & 0.0250 & 0.0232 & 0.0329 & 1.42x \\
0.030 & 33,728    & 0.0150 & 0.0144 & 0.0178 & 1.23x \\
\bottomrule
\end{tabular}
\end{table}

\subsection{Priority-Queue Comparison}

\begin{figure}[t!]
    \centering
    \includegraphics[width=\linewidth]{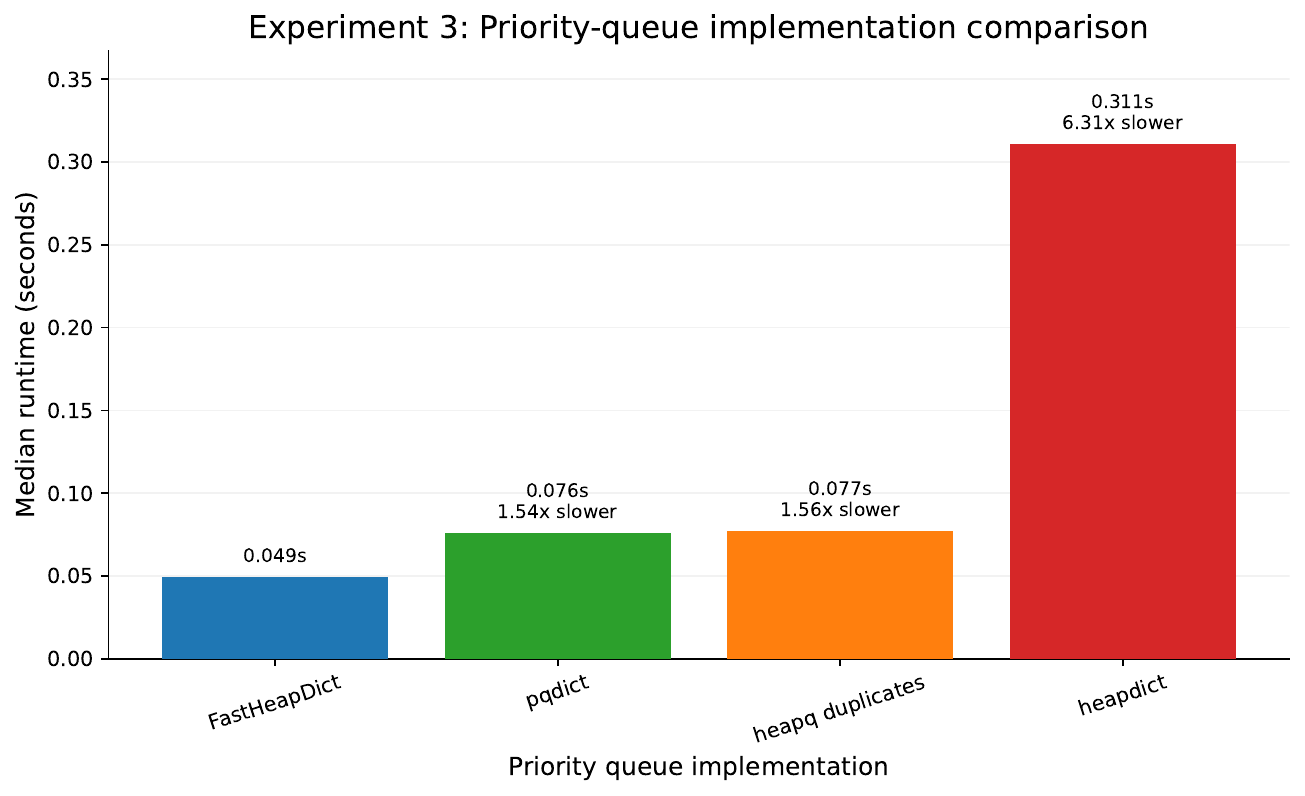}
    \caption{Python Experiment~3: comparison of Python priority-queue implementations at $V=1500$ and density ratio $0.10$.}
    \label{fig:python_exp3}
\end{figure}

Table~\ref{tab:python_pq} compares FastHeapDict with three widely used Python priority-queue implementations: duplicate-entry \texttt{heapq}, \texttt{pqdict}, and \texttt{heapdict}, while Fig.~\ref{fig:python_exp3} visualizes the corresponding runtime comparison. FastHeapDict achieves the lowest running time among all evaluated methods. Although both \texttt{pqdict} and \texttt{heapdict} support indexed priority updates, their practical overhead differs substantially. On this workload, \texttt{pqdict} performs comparably to duplicate-entry \texttt{heapq}, whereas \texttt{heapdict} is noticeably slower.

\begin{table}[t]
\caption{Python priority-queue comparison for $V=1500$ and density ratio $0.10$.}
\label{tab:python_pq}
\centering
\scriptsize
\begin{tabular}{l r r r r r}
\toprule
Method & Median (s) & Updates & Pops & Stale & Max \\
\midrule
FastHeapDict & \textbf{0.0493} & 112,049 & \textbf{899} & \textbf{0} & 898 \\
pqdict       & 0.0758          & 112,049 & \textbf{899} & \textbf{0} & 898 \\
heapq dup.   & 0.0772          & 112,048 & 111,001       & 110,103    & 111,270 \\
heapdict     & 0.3109          & 112,049 & \textbf{899} & \textbf{0} & \textbf{897} \\
\bottomrule
\end{tabular}
\end{table}

\subsection{Randomized Dense Stress Graphs}

\begin{figure}[t]
    \centering
    \includegraphics[width=\linewidth]{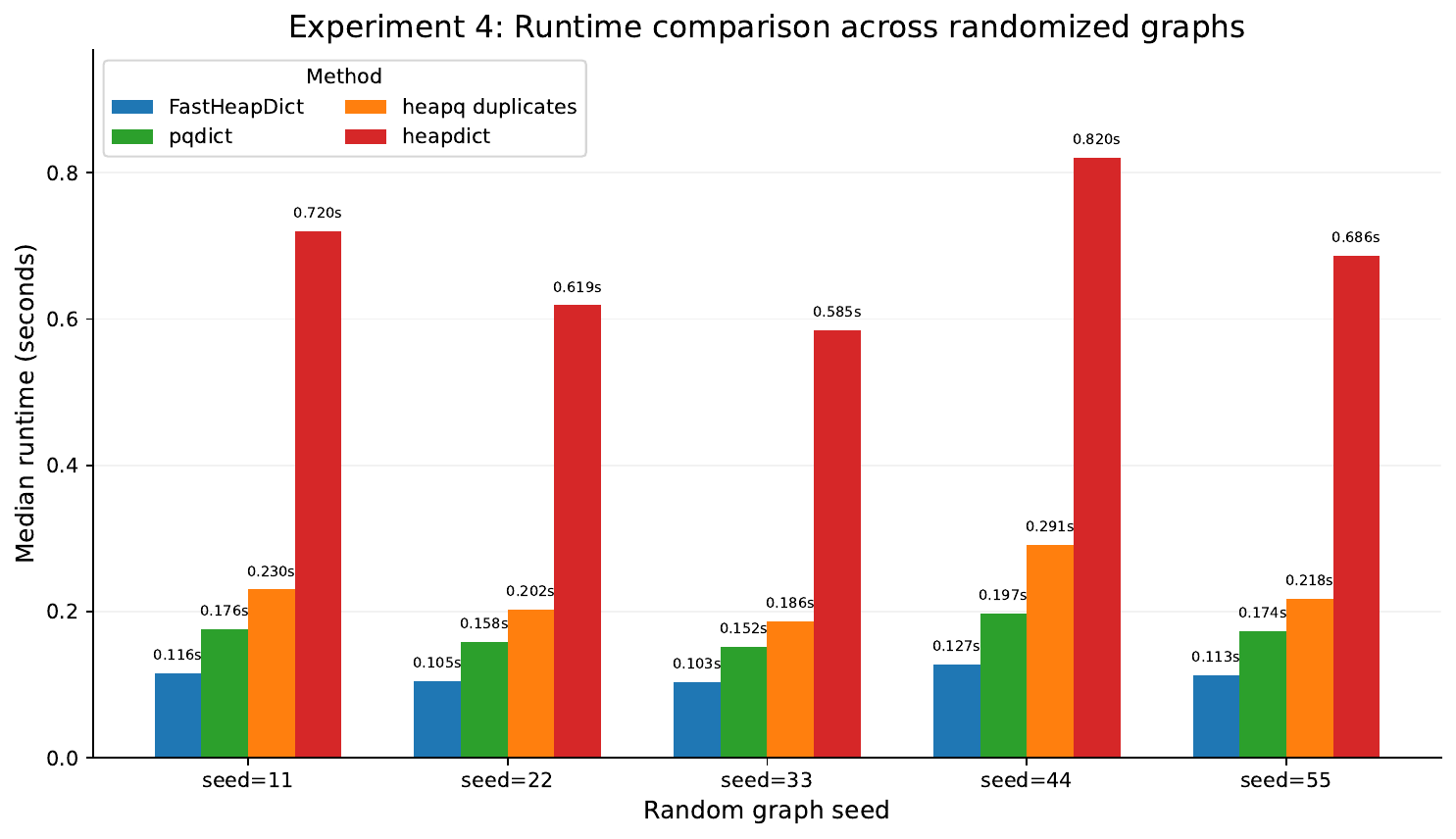}
    \caption{Python Experiment~4: running times on randomized dense stress graphs.}
    \label{fig:python_exp4}
\end{figure}

Experiment~4 evaluates five independently generated randomized dense stress graphs in order to verify that the observed behavior is not specific to a single deterministic construction. Table~\ref{tab:python_random} shows that FastHeapDict consistently outperforms duplicate-entry \texttt{heapq} for every random seed, and Fig.~\ref{fig:python_exp4} visualizes the running times across the randomized graph instances. The measured speedups range from $\mathbf{1.80\times}$ to $\mathbf{2.29\times}$. Throughout these experiments, FastHeapDict maintains exactly one live heap entry per vertex, whereas the duplicate-entry implementation processes hundreds of thousands of stale removals before termination.

\begin{table}[t]
\caption{Python randomized dense stress graphs at $V=1500$ and density ratio $0.25$.}
\label{tab:python_random}
\centering
\scriptsize
\begin{tabular}{l r r r r}
\toprule
Seed & Active sinks & FastHeapDict (s) & heapq dup. (s) & Speedup \\
\midrule
11 & 310 & 0.1156 & 0.2302 & 1.99x \\
22 & 271 & 0.1048 & 0.2022 & 1.93x \\
33 & 251 & 0.1032 & 0.1862 & 1.80x \\
44 & 372 & 0.1275 & 0.2915 & \textbf{2.29x} \\
55 & 293 & 0.1128 & 0.2177 & 1.93x \\
\bottomrule
\end{tabular}
\end{table}

\FloatBarrier
\section{C++20 Results}
\label{sec:cpp-results}

\subsection{Scaling with the Number of Vertices}

\begin{figure}[t!]
    \centering
    \includegraphics[width=\linewidth]{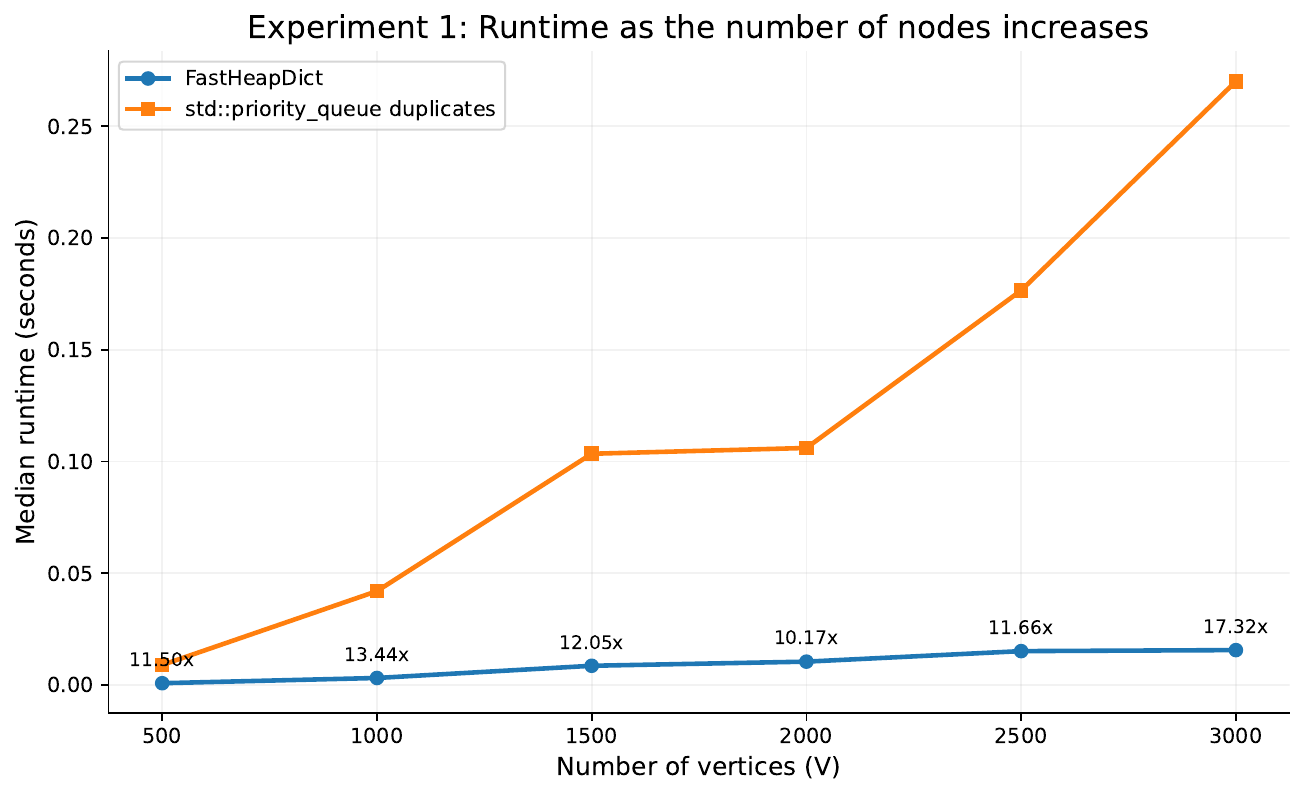}
    \caption{C++20 Experiment~1: running time as the number of vertices increases while the density ratio is fixed at $0.25$.}
    \label{fig:cpp_exp1}
\end{figure}

Table~\ref{tab:cpp_scaling} summarizes the median running times obtained in Experiment~1 for the C++20 implementation, and Fig.~\ref{fig:cpp_exp1} visualizes the corresponding runtime trend. The graph density ratio is fixed at $\rho=0.25$, while the number of vertices increases from 500 to 3000. FastHeapDict consistently outperforms the duplicate-entry \texttt{std::priority\_queue} implementation across the entire range of graph sizes. The measured speedup is at least $\mathbf{10.17\times}$ across all tested graph sizes and reaches $\mathbf{17.32\times}$ for the largest graph, demonstrating that the practical advantage of the proposed design remains substantial as the workload increases.

\begin{table}[H]
\caption{C++20 median running time as the graph size increases at density ratio $0.25$.}
\label{tab:cpp_scaling}
\centering
\scriptsize
\setlength{\tabcolsep}{3.5pt}
\begin{tabular}{l r r r r r}
\toprule
$V$ & $E$ & Density & FastHeapDict (s) & std::pq dup. (s) & Speedup \\
\midrule
500 & 31,188 & 0.250 & 0.000778 & 0.008943 & 11.50x \\
1,000 & 124,875 & 0.250 & 0.003122 & 0.041968 & 13.44x \\
1,500 & 281,062 & 0.250 & 0.008582 & 0.103429 & 12.05x \\
2,000 & 499,750 & 0.250 & 0.010428 & 0.106022 & 10.17x \\
2,500 & 780,938 & 0.250 & 0.015138 & 0.176501 & 11.66x \\
3,000 & 1,124,625 & 0.250 & 0.015592 & 0.270049 & \textbf{17.32x} \\
\bottomrule
\end{tabular}
\end{table}

\subsection{Changing Graph Density}

\begin{figure}[t]
    \centering
    \includegraphics[width=\linewidth]{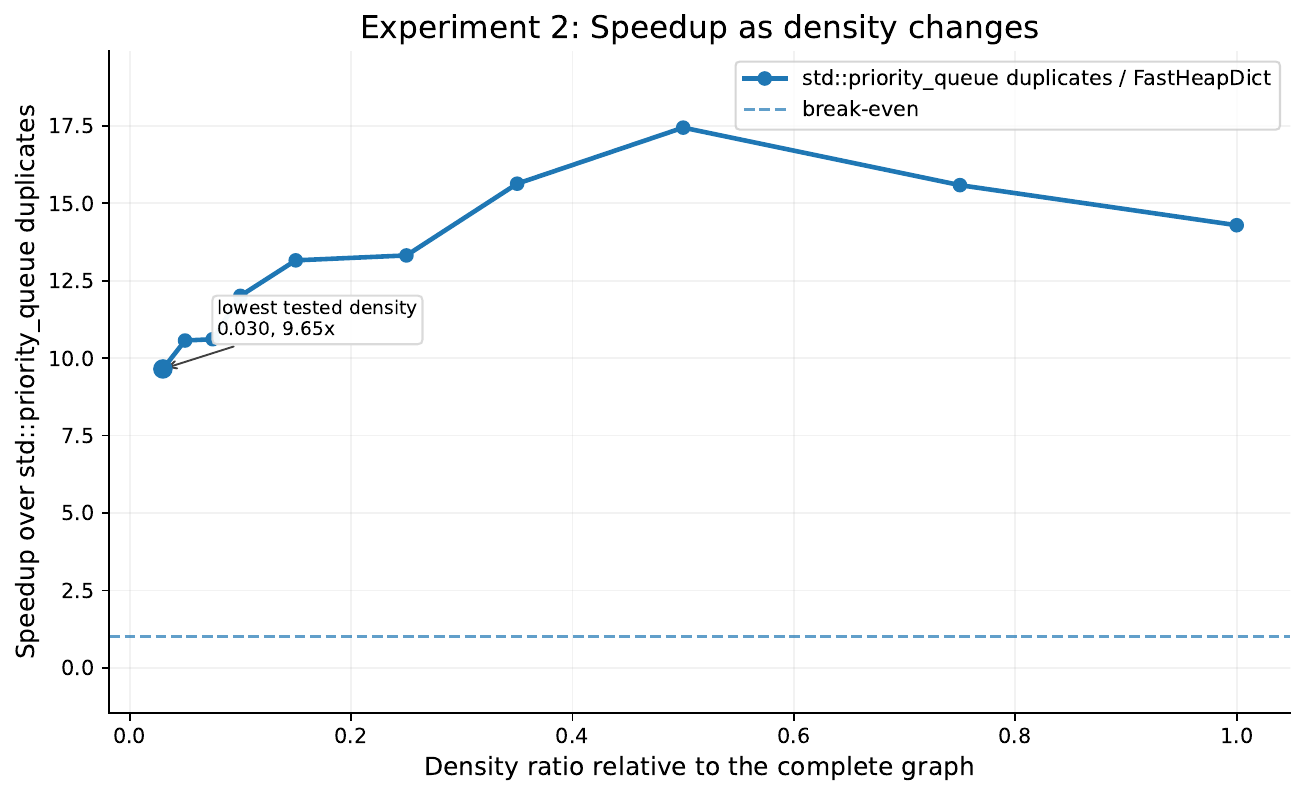}
    \caption{C++20 Experiment~2: speedup of FastHeapDict relative to duplicate-entry \texttt{std::priority\_queue} as graph density varies.}
    \label{fig:cpp_exp2}
\end{figure}

Experiment~2 fixes the graph size at $V=1500$ while varying the density ratio from $0.03$ to $1.0$. Table~\ref{tab:cpp_density} reports the corresponding median running times, and Fig.~\ref{fig:cpp_exp2} shows the resulting speedup trend over the density sweep. This density sweep follows the two-method design used in the corresponding Python experiment and compares FastHeapDict against the duplicate-entry \texttt{std::priority\_queue} baseline. FastHeapDict remains consistently faster throughout the entire range of tested densities. Measured speedups vary from $\mathbf{9.65\times}$ to $\mathbf{17.44\times}$, with the largest speedup occurring at density ratio $0.50$. Unlike the Python implementation, where the performance gap narrows noticeably for sparse graphs, the C++ implementation maintains a substantial advantage even at the lowest tested density.

\begin{table}[t]
\caption{C++20 median running time as graph density varies at $V=1500$.}
\label{tab:cpp_density}
\centering
\scriptsize
\setlength{\tabcolsep}{3.0pt}
\begin{tabular}{l r r r r r}
\toprule
Density & $E$ & $\alpha$ & FastHeapDict (s) & std::pq dup. (s) & Speedup \\
\midrule
1.000 & 1,124,250 & 0.4997 & 0.008486 & 0.121276 & 14.29x \\
0.750 & 843,188 & 0.3748 & 0.008039 & 0.125288 & 15.59x \\
0.500 & 562,125 & 0.2498 & 0.006937 & 0.121001 & \textbf{17.44x} \\
0.350 & 393,488 & 0.1749 & 0.005080 & 0.079402 & 15.63x \\
0.250 & 281,062 & 0.1249 & 0.004099 & 0.054588 & 13.32x \\
0.150 & 168,638 & 0.0750 & 0.002380 & 0.031310 & 13.16x \\
0.100 & 112,425 & 0.0500 & 0.001884 & 0.022632 & 12.01x \\
0.075 & 84,319 & 0.0375 & 0.001447 & 0.015353 & 10.61x \\
0.050 & 56,212 & 0.0250 & 0.000920 & 0.009725 & 10.57x \\
0.030 & 33,728 & 0.0150 & 0.000585 & 0.005647 & 9.65x \\
\bottomrule
\end{tabular}
\end{table}

\begin{table*}[t!]
\caption{C++20 priority-queue comparison for $V=1500$ and density ratio $0.10$.}
\label{tab:cpp_pq}
\centering
\small
\begin{tabular}{l r r r r r}
\toprule
Method & Median (s) & Updates & Pops & Stale pops & Max heap \\
\midrule
FastHeapDict & \textbf{0.001842} & 112,033 & \textbf{899} & \textbf{0} & \textbf{897} \\
std\_set\_decrease\_key & 0.010513 & 112,033 & \textbf{899} & \textbf{0} & \textbf{897} \\
std\_priority\_queue\_duplicates & 0.019360 & 112,032 & 111,001 & 110,103 & 111,243 \\
std\_multiset\_duplicates & 0.021280 & 112,032 & 111,001 & 110,103 & 111,243 \\
\bottomrule
\end{tabular}
\end{table*}

\subsection{Priority-Queue Comparison}

\begin{figure}[!t]
    \centering
    \includegraphics[width=\linewidth]{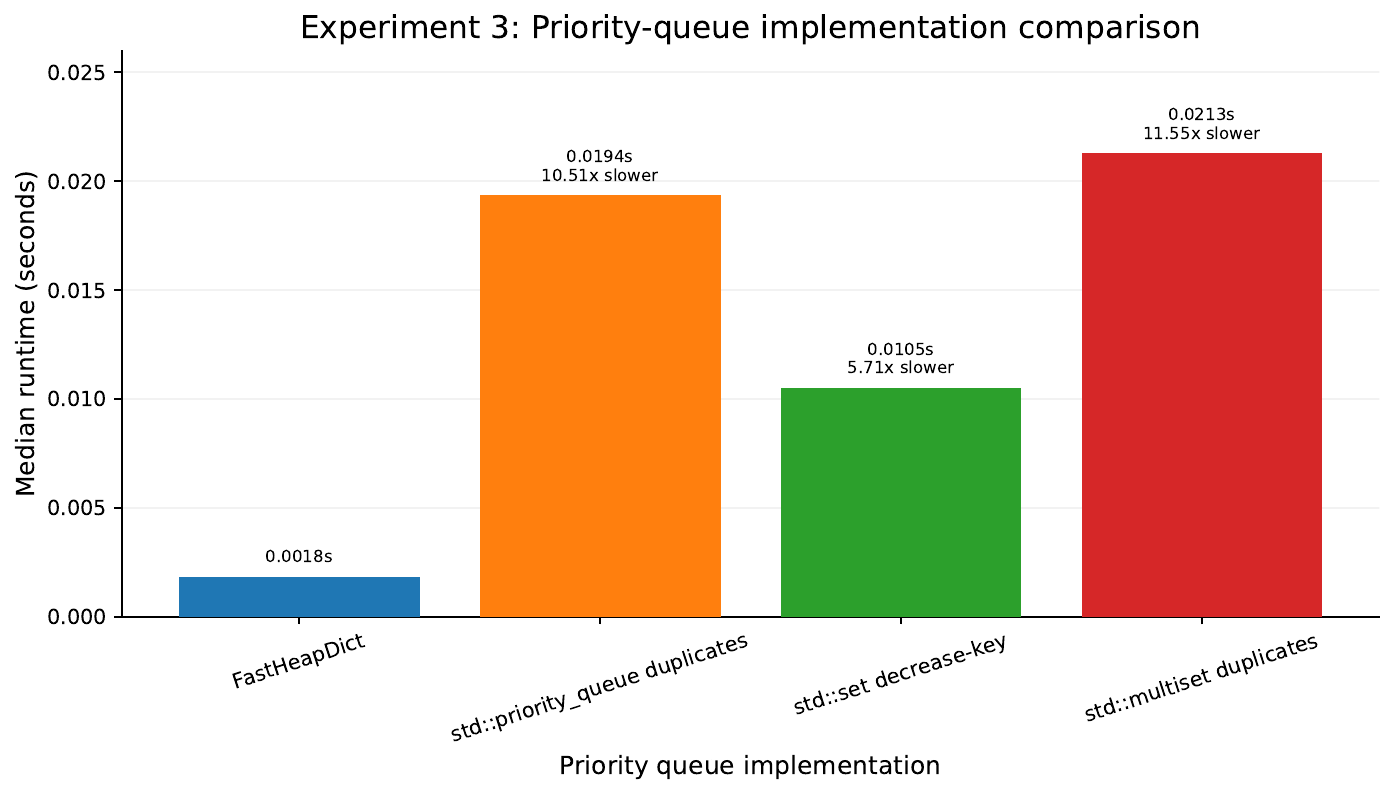}
    \caption{C++20 Experiment 3: priority-queue comparison at $V=1500$ and density ratio $0.10$.}
    \label{fig:cpp_exp3}
\end{figure}

Table~\ref{tab:cpp_pq} compares FastHeapDict with three commonly used C++ priority-queue implementations: duplicate-entry \texttt{std::priority\_queue}, tree-based \texttt{std::set} with erase-and-reinsert updates, and duplicate-entry \texttt{std::multiset}; Fig.~\ref{fig:cpp_exp3} visualizes the corresponding runtime comparison. FastHeapDict achieves the lowest running time among all evaluated methods. Although the \texttt{std::set} implementation maintains a single live entry for each vertex, its balanced-tree operations incur substantially larger constant factors than the binary-heap-based design. Both duplicate-entry baselines also perform a large number of stale removals, resulting in considerably larger queue sizes during execution.

\begin{table*}[t]
\caption{C++20 randomized dense stress graphs at $V=1500$ and density ratio $0.25$.}
\label{tab:cpp_random}
\centering
\scriptsize
\setlength{\tabcolsep}{3.0pt}
\begin{tabular}{l r r r r r r r}
\toprule
Seed & Active sinks & $E$ & $\alpha$ & FastHeapDict (s) & std::pq dup. (s) & std::set (s) & std::multiset dup. (s) \\
\midrule
11 & 308 & 281,062 & 0.1249 & \textbf{0.003304} & 0.055201 & 0.057951 & 0.113341 \\
22 & 270 & 281,062 & 0.1249 & \textbf{0.002832} & 0.049830 & 0.045843 & 0.095882 \\
33 & 303 & 281,062 & 0.1249 & \textbf{0.003855} & 0.050241 & 0.051738 & 0.117295 \\
44 & 265 & 281,062 & 0.1249 & \textbf{0.003021} & 0.042370 & 0.043169 & 0.097723 \\
55 & 311 & 281,062 & 0.1249 & \textbf{0.003348} & 0.056428 & 0.058424 & 0.110485 \\
\bottomrule
\end{tabular}
\end{table*}

\subsection{Randomized Dense Stress Graphs}

\begin{figure}[!t]
    \centering
    \includegraphics[width=\linewidth]{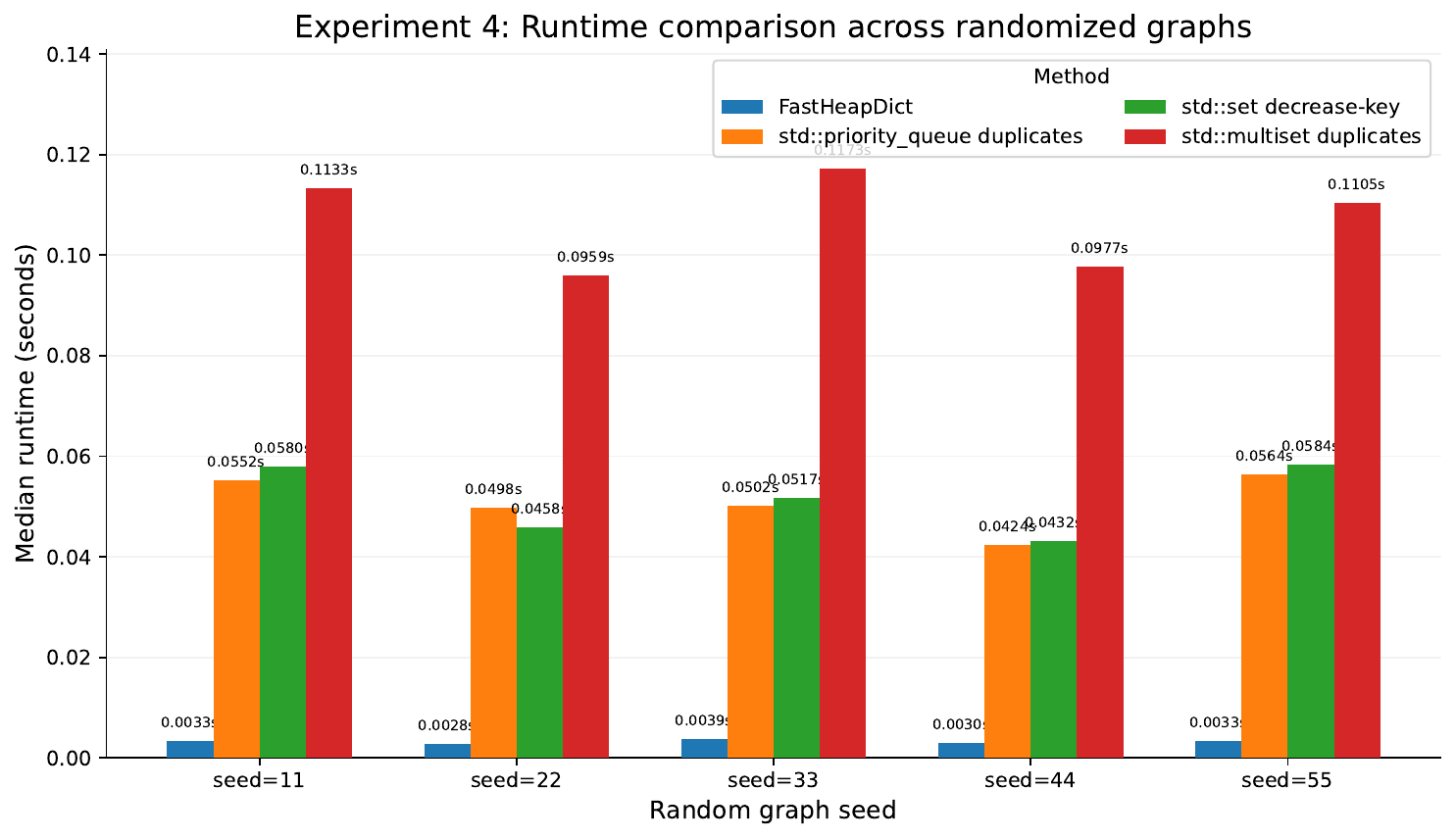}
    \caption{C++20 Experiment 4: randomized dense stress-graph running times by method.}
    \label{fig:cpp_exp4}
\end{figure}

Experiment~4 evaluates five independently generated randomized dense stress graphs using the C++20 implementation. Table~\ref{tab:cpp_random} reports all four C++ priority-queue methods for each seed, and Fig.~\ref{fig:cpp_exp4} visualizes the running times across the randomized graph instances. FastHeapDict is the fastest method on every randomized graph. Relative to duplicate-entry \texttt{std::priority\_queue}, the observed FastHeapDict speedups range from $\mathbf{13.03\times}$ to $\mathbf{17.60\times}$. The \texttt{std::set} decrease-key implementation is close to the duplicate-entry heap baseline on these randomized graphs, while \texttt{std::multiset} is consistently slower. This comparison shows that the advantage is not only due to avoiding stale duplicate entries, but also to the lower constant-factor overhead of the vector-backed permanent-slot heap.

\FloatBarrier
\section{Discussion}
\label{sec:discussion}

The experimental results consistently support the central design hypothesis underlying FastHeapDict. By maintaining exactly one live heap entry for each key and replacing repeated key-based position updates with permanent slot indexing, the proposed implementation reduces the practical overhead associated with decrease-key operations while preserving the asymptotic complexity of a conventional indexed binary heap.

The observed speedups therefore arise from reducing implementation overhead rather than from introducing a new heap algorithm or improving asymptotic complexity.

One observation common to both implementations is that the benefit of FastHeapDict increases as duplicate-entry overhead becomes more significant. In the constructed dense stress graphs, repeated edge relaxations generate large numbers of obsolete heap entries in duplicate-entry implementations. Every stale entry eventually requires an additional heap removal, increasing both the total number of heap operations and the effective queue size during execution. Because FastHeapDict updates priorities in place, this overhead is avoided entirely.

The magnitude of the improvement differs between Python and C++20. In Python, FastHeapDict consistently outperforms duplicate-entry \texttt{heapq} by approximately a factor of two on the evaluated workloads. This result is noteworthy because \texttt{heapq} is a mature and highly optimized standard-library implementation, while FastHeapDict is implemented entirely in pure Python. The C++20 implementation exhibits substantially larger speedups. In addition to outperforming duplicate-entry \texttt{std::priority\_queue}, FastHeapDict also consistently exceeds the performance of the tree-based \texttt{std::set} implementation, indicating that an indexed binary heap with efficient position maintenance can provide lower practical overhead than balanced-tree alternatives for decrease-key-intensive workloads.

The randomized dense stress graphs strengthen this conclusion by demonstrating that the observed behavior is not tied to a single handcrafted graph instance. Although individual graph structures vary across random seeds, FastHeapDict consistently maintains its performance advantage. This suggests that the improvement arises from the underlying implementation strategy rather than from properties of one particular benchmark graph.

At the same time, the present results should be interpreted within the scope of the evaluated workloads. The benchmark families were intentionally designed to emphasize repeated priority updates in order to isolate the practical cost of duplicate-entry processing. Consequently, the reported speedups should not be interpreted as universal improvements for every graph or every application. Instead, they demonstrate that permanent slot indexing provides a practical implementation strategy for indexed priority queues whose benefits become increasingly apparent as decrease-key operations become more frequent.
\section{Limitations}
\label{sec:limitations}

The present study has several limitations that should be considered when interpreting the reported results.

First, the benchmark graphs are synthetic and were intentionally designed to emphasize workloads with frequent decrease-key operations. Although this construction provides a controlled setting for evaluating the proposed implementation, it does not capture the full diversity of real-world graph families such as road networks, social networks, communication graphs, or finite-element meshes.

Second, the experimental evaluation focuses on positive-weight single-source shortest-path computations using Dijkstra's algorithm. Other application domains, including multi-source shortest paths, all-pairs shortest paths, dynamic graphs, or algorithms such as Prim's minimum spanning tree algorithm and A* search, may exhibit different performance characteristics and warrant separate investigation.

Third, all experiments were performed on a single hardware and software platform. Although both the Python and C++20 implementations consistently demonstrate the same qualitative behavior, absolute running times and constant factors may vary across processors, compilers, optimization settings, memory allocators, and operating systems.

Finally, FastHeapDict does not alter the asymptotic complexity of indexed binary heaps. Its contribution is implementation-oriented rather than theoretical: by introducing permanent slot indexing, it reduces repeated key-based hash-table accesses during heap maintenance while preserving the standard time complexity of indexed binary heaps. Broader evaluation across additional graph families, algorithms, and computing environments will help further characterize the practical applicability of the proposed design.
\FloatBarrier
\section{Conclusion}
\label{sec:conclusion}

This paper introduced \emph{FastHeapDict}, an indexed binary heap based on the principle of permanent slot indexing. Rather than repeatedly updating heap positions through key-based hash-table operations during heap maintenance, FastHeapDict assigns each key a stable integer slot that is reused throughout its lifetime. This design preserves the standard asymptotic complexity of indexed binary heaps while reducing the practical overhead associated with frequent decrease-key operations.

The proposed design was evaluated through independent Python and C++20 implementations using Dijkstra's algorithm as the primary experimental workload. Across all evaluated benchmark families, including deterministic and randomized dense stress graphs, FastHeapDict consistently outperformed duplicate-entry priority queues. The Python implementation also exceeded the performance of existing indexed priority-queue libraries, while the C++20 implementation demonstrated substantial constant-factor improvements over both duplicate-entry and tree-based standard-library implementations.

The contribution of this work is implementation-oriented rather than theoretical. FastHeapDict does not introduce a new priority-queue complexity bound; instead, it demonstrates that careful engineering of indexed priority queues can produce substantial practical improvements while preserving the simplicity and theoretical guarantees of binary heaps. We hope that the proposed design will serve as a useful reference for future implementations of decrease-key-intensive graph algorithms and motivate further investigation of practical priority-queue engineering across a broader range of algorithms, workloads, and computing environments.

\section*{Reproducibility Notes}

All benchmark artifacts required to reproduce the reported
results are publicly available in the project's GitHub repository.
These include the Python and C++20 implementations, benchmark
scripts, CSV benchmark outputs, plotting scripts, and the
generated figures used throughout the evaluation.
The experimental hardware and software environment is described
in Section~\ref{sec:system_configuration}.
\section*{Code and Reproducibility}
\label{sec:code-availability}

The complete source code, benchmark scripts, raw experimental data, plotting utilities, generated figures, and detailed reproduction notes are publicly available in the project repository: \url{https://github.com/hojjatgoudarzi/FastHeapDict}. For convenient reuse, the Python implementation is additionally distributed through the Python Package Index and can be installed with \texttt{pip install fastheapdict}.

\bibliographystyle{IEEEtran}
\bibliography{references}

\end{document}